\magnification=1200
\baselineskip 12pt
\def\picture #1 by #2 (#3){
  \vbox to #2{
    \hrule width #1 height 0pt depth 0pt
    \vfill
    \special{picture #3}
    }
  }

\def\scaledpicture #1 by #2 (#3 scaled #4){{
  \dimen0=#1 \dimen1=#2
  \divide\dimen0 by 1000 \multiply\dimen0 by #4
  \divide\dimen1 by 1000 \multiply\dimen1 by #4
  \picture \dimen0 by \dimen1 (#3 scaled #4)}
  }


\vglue2cm 

\bf

\centerline{BIRTH OF RESONANCES IN THE SPIN--ORBIT}

\medskip

\centerline{PROBLEM OF CELESTIAL MECHANICS}

\rm

\vskip.2in 

\centerline{\bf Alessandra Celletti$^{(1)}$ and Luigi Chierchia$^{(2)}$}

\vglue1cm

(1) Dipartimento di Matematica Pura
e Applicata, Universit\`a di L'Aquila, 
Via Vetoio - I--67010 L'Aquila (Italy) 

e--mail: alessandra.celletti@aquila.infn.it

\vskip.2in 

(2) Dipartimento di Matematica, Universit\`a Roma Tre, 
Largo San Leonardo Murialdo 1, I--00146 Roma (Italy)

e--mail: luigi@matrm3.mat.uniroma3.it

\vglue2cm\noindent
\bf ABSTRACT. \rm The behaviour of resonances in the spin--orbit coupling 
in Celestial Mechanics is investigated. We introduce a Hamiltonian 
nearly--integrable model describing an approximation of the spin--orbit 
interaction. A parametric representation of periodic orbits is presented. 
We provide explicit formulae to compute the Taylor series
expansion in the perturbing parameter of the function describing this 
parametrization. Then we compute approximately the radius of convergence providing 
an indication of the stability of the periodic orbit. This quantity is used to
describe the different probabilities of capture into resonance. In particular,
we notice that for low values of the orbital eccentricity the only significative
resonance is the synchronous one. Higher order resonances (including 1:2, 3:2,
2:1) appear only as the orbital eccentricity is increased.

\vglue2cm\noindent

\vfill\eject 

\noindent 
\bf \S1. INTRODUCTION\rm 

\vskip.1in\noindent
The dynamics of coupling between revolutional and rotational motions of a
satellite is investigated. In particular, we consider an oblate satellite moving
on a Keplerian orbit around a central planet and rotating about an internal
spin--axis (see Celletti 1990, Celletti 1994, Celletti and Falcolini 1992, 
Celletti and Chierchia 1998, Goldreich and Peale 1966, 1970, Henrard 1985, Murdock
1978, Peale 1973, Wisdom 1987 and references therein for related papers 
on this subject). 
Whenever the ratio between the revolutional and rotational periods
is rational, say ${{T_{rev}}\over{T_{rot}}}={p\over q}$ for some positive integers $p$,
$q$, a spin--orbit resonance (of order $p:q$) is encountered. Astronomical
observations show that most of the evolved satellites or planets of the solar
system are trapped in a 1:1 or "synchronous" spin--orbit resonance. The most
familiar example is provided by the Moon; due to the equality of the periods of
rotation and revolution, the Moon always points the same face to the Earth. The
only known exception is provided by Mercury, which moves in a  3:2 resonance,
namely after two revolutions around the Sun, Mercury makes three rotations about
its spin--axis. An intriguing question is the explanation of the different
occurrence between \sl plenty \rm resonances (like the 1:1 or 3:2) and \sl empty \rm 
resonances (corresponding to any other different ratio of the periods). In this paper
we try to address this problem as follows. We introduce a mathematical non--autonomous,
one--dimensional model describing  a physically relevant approximation of the
spin--orbit problem (see \S2). The equation of motion takes the form 
$$
\ddot x\ -\ \varepsilon\ f_x(x,t)\ =\ 0\ , 
$$
for a suitable function $f$ depending also on the orbital eccentricity $e$ of
the Keplerian orbit; the parameter $\varepsilon$ is proportional to the
equatorial oblateness coefficient of the satellite. 

\vskip.1in\noindent
In \S3 we introduce a parametric representation of a periodic orbit with
frequency $p\over q$ as 
$$\eqalign{ 
x&=\theta+u(\theta)\cr
y&=v(\theta)\qquad\qquad\theta\in{\bf T}\equiv {\bf R}/(2\pi{\bf Z})\ ,\cr} 
$$
where $u$, $v$ are suitable functions depending analytically on $\varepsilon$ 
and $\theta(t)=\theta(0)+{p\over q}t$. 
We provide explicit formulae for the computation of the coefficients of the
Taylor series expansion around $\varepsilon=0$ of the function $u$, say 
$u(\theta)=\sum_{j=1}^\infty u_j(\theta)\varepsilon^j$. 
Finally, we compute an approximation of the radius of convergence 
$$
\varepsilon_\rho({p\over q})\ \equiv\ \lim_{j\to\infty} {1\over
{(u_j(\theta_0))^{1\over j}}} 
$$
(for a fixed $\theta=\theta_0$). This quantity provides a measure of the
regularity of the periodic orbit and is used to provide a qualitative argument
toward a higher capture probability of the synchronous resonance. 
We remark that a recent method based on a quantitative version of the Implicit
Function Theorem has been implemented in (Celletti and Chierchia, 1998) to construct
Birkhoff periodic orbits.

\vskip.2in\noindent
The results and conclusions are presented in \S4. The graphs of 
$\varepsilon_\rho({p\over q})$ vs. ${p\over q}$ show that for the Moon and
Mercury, the 1:1 and 3:2 resonances respectively seem to be the most suitable
ending--states. Moreover we explore the \sl birth \rm of resonances as the
orbital eccentricity is increased. In particular, for low values of $e$ the
synchronous resonance is the only one present. As the eccentricity 
increases, higher order resonances do appear. We relate these results to the
evolutionary history of the satellites. 

\vskip.2in\noindent
We remark that an exhaustive explanation of this scenario requires the
consideration of dissipative forces and their role in stabilizing the
resonances. We demand this problem to a future study.

\vglue2cm 

\noindent 
\bf \S2. THE SPIN--ORBIT MODEL\rm 

\vskip.1in\noindent
\vskip.1in\noindent
We briefly introduce a mathematical model describing the 
``spin--orbit" interaction in Celestial Mechanics as follows. 

\medskip\noindent
Let $S$ be a triaxial ellipsoidal satellite orbiting around a central planet $P$. Let 
$T_{rev}$ and $T_{rot}$ be the periods of revolution of the satellite around $P$
and the period of rotation about an internal spin--axis. 
A \sl $p:q$ spin--orbit resonance \rm occurs whenever 
$$ {{T_{rev}}\over {T_{rot}}}\ =\ {p\over q}\ ,\qquad\qquad {\rm for}\ p,q\in {\bf
N},\ q\not=0\ . 
$$ 
In particular, when $p=q=1$ we speak of 1:1 or \sl synchronous \rm spin--orbit 
resonance, which implies that the satellite always points the same face to the host planet.
In the solar system most of the \sl evolved \rm satellites or planets 
(like, e.g., the Moon) are trapped in a 1:1 resonance (Astronomical Almanac, 1997). 
The only exception is provided by Mercury which is observed in a nearly 3:2 resonance. 

We introduce a mathematical model describing the spin--orbit interaction. In 
particular we assume that 

\noindent 
$i)$ the center of mass of the  satellite moves on a Keplerian orbit around  $P$ with
semimajor axis $a$ and eccentricity $e$ (i.e., we neglect secular perturbations on the  orbital
parameters); 

\noindent 
$ii)$ the spin--axis is perpendicular to the orbit plane (i.e., we neglect the 
``obliquity"); 

\noindent 
$iii)$ the spin--axis coincides with the shortest physical axis (i.e., the  axis whose
moment of inertia is largest); 

\noindent 
$iv)$ dissipative effects as well as perturbations due to other planets or  satellites
are neglected.

\medskip\noindent
Let $A<B<C$ be the principal moments of inertia of the satellite. We denote by $r$ and
$f$, respectively, the instantaneous orbital radius and the true anomaly of the
Keplerian  orbit.  Finally let $x$ be the angle between the longest axis of the
ellipsoid and the periapsis  line (see
Figure 1). The equation of motion governing the spin--orbit model 
under the assumptions $i)-iv)$ may be derived from the standard
Euler's equations for rigid body. In normalized units (i.e. assuming that the mean motion is one, 
$2\pi/T_{rev}=1$) we obtain:  
$$
\ddot x\ +\  \varepsilon ({1\over r})^3\ \sin (2x-2f)\ =\ 0\ , 
\eqno{(2.1)} 
$$  
where $\varepsilon\equiv{3\over 2}{{B-A}\over C}$ is proportional to the equatorial 
oblateness coefficient ${{B-A}\over C}$ (and the dot denotes time differentiation). Notice that
$(2.1)$ is trivially integrated when $A=B$ or in the case of zero orbital eccentricity. 

\medskip\noindent 
A ``$p:q$ periodic orbit" (or ``Birkhoff periodic orbit with rotation number $p/q$") is a
solution of $(2.1)$ such that 
$$ 
x(t+2\pi q)\ =\ x(t)+2\pi p\ ;  
$$ 
the above relation implies that after $q$ orbital revolutions around the central body 
the satellite makes $p$ rotations about the spin--axis.

\vskip0.5cm\noindent 
Due to assumption $i)$, the quantities $r$ and $f$ are known Keplerian functions of
the time; therefore we can expand $(2.1)$ in Fourier series as 
$$
\ddot x\ +\ \varepsilon\ \sum\limits_{m\not=0, m=-\infty}^{\infty}  W({m\over 2},e)\
\sin(2x-mt)\ =\ 0\ , 
\eqno{(2.2)} 
$$  
where the coefficients $W({m\over 2},e)$ decay as powers of the  orbital
eccentricity as $W({m\over 2},e)\propto e^{|m-2|}$ (see Cayley 1859, for explicit
expressions). 

\medskip\noindent 
A further simplification of the model is performed as follows. 
According to assumption $iv)$,  we
neglected dissipative forces and any gravitational attraction beside that of the
central planet. The most important contribution comes from the non--rigidity of the
satellite,  which provokes a tidal torque due to internal friction. 
Since the magnitude of the dissipative effects is
small compared to the gravitational  term, we simplify $(2.2)$ retaining
only those terms which are of the same order or bigger than the average
effect of the  tidal torque. Finally we consider an equation of the form 
$$
\ddot x\ +\ \varepsilon\ \sum_{m\not=0,m=N_1}^{N_2}\ \tilde W({m\over 2},e)\ 
\sin(2x-mt)\ =\ 0\ ,
$$  
where $N_1$ and $N_2$ are suitable integers (depending on the structural and orbital 
properties of the satellite), while $\tilde W({m\over 2},e)$ are truncations of the coefficients
$W({m\over 2},e)$ (which are power series in e).  

\medskip\noindent
According to the above discussion, in the case of the Moon--Earth system we consider the 
following equation of motion: 
$$\eqalign{
\ddot x\ &+\ \varepsilon\ [(-{e\over 2}+{{e^3}\over {16}})\ \sin(2x-t)\ +\cr
&+(1-{5\over 2}e^2+{{13}\over {16}}e^4)\
\sin(2x-2t)\ +\ ({7\over 2} e-{{123}\over {16}}e^3)\ \sin(2x-3t)\ +\cr &+({{17}\over
{2}}e^2-{{115}\over {6}}e^4)\ \sin(2x-4t)\ +\ ({{845}\over {48}}e^3-{{32525}\over
{768}}e^5)\ \sin(2x-5t)\ + \cr &+{{533}\over{16}}e^4\sin(2x-6t)\ +\
{{228347}\over{3840}}e^5\ \sin(2x-7t)\ ]\ =\ 0\ ,\cr}
\eqno(2.3)
$$
having taken $N_1=1$ and $N_2=7$ in $(2.2)$. In what follows we shall always assume 
that the leading equation is provided by $(2.3)$, though in principle we should consider 
different truncation orders ($N_1$, $N_2$) depending on the intrinsic parameters of
the  satellite. For example, in the Mercury--Sun case the above
criterion would suggest to take $N_1=-17$ and $N_2=6$. However we checked 
that our results do not change significantly as new terms are added to eq. $(2.3)$. 
Precisely, more terms alter the outputs for \sl any \rm initial condition, 
provoking a global rescaling of the results, without conditioning the
qualitative conclusions we shall draw about the behaviour of the resonances. 
As an illustration we 
report in Figure 2 the Poincar\'e map around the synchronous resonance 
for $e=0.004$, $\varepsilon=0.2$. The resonance is surrounded by \sl librational
\rm surfaces, whose amplitude increases as the \sl chaotic separatrix \rm is
approached. Outside this region \sl rotational \rm surfaces can be found.

\vglue2cm 

\noindent 
\bf \S3. PARAMETRIC REPRESENTATION OF PERIODIC ORBITS\rm 

\vskip.1in\noindent
In this section we provide a parametric representation of the solution associated 
to a periodic orbit with frequency $\omega={p\over q}$. 

\noindent 
We rewrite eq. (2.3) in compact form as 
$$
\ddot x\ -\ \varepsilon f_x(x,t)\ =\ 0\ , 
$$
namely 
$$\eqalign{ 
\dot x&= y \cr
\dot y&=\varepsilon f_x(x,t)\ .\cr} 
\eqno{(S)} 
$$
A periodic orbit with frequency $\omega={p\over q}$ is defined parametrically by
the set of equations 
$$\eqalign{ 
x&=\theta+u(\theta)\cr
y&={p\over q}+Du(\theta)\qquad\qquad \theta\in{\bf T}\ , \cr} 
\eqno{(3.2)}
$$
where $u=u(\theta)$ is a suitable continuous function, depending analytically 
on $\varepsilon$, with the property that the flow in the $\theta$--coordinate is 
linear, i.e. $\theta\rightarrow\theta+{p\over q}t$ after a time $t$; 
the operator $D$ acts on a function $u=u(\theta)$ as 
$$
Du(\theta)=\omega\ {{d u(\theta)}\over{d\theta}}\ =\ {p\over q}\ 
{{d u(\theta)}\over{d\theta}}\ . 
$$

\vskip.2in
\noindent 
Inserting $(3.2)$ in the system $(3.1)$ one obtains the second order differential
equation 
$$
D^2u(\theta)-\varepsilon f_x(\theta+u(\theta),{\theta\over\omega})\ =\ 0\ . 
\eqno{(3.3)} 
$$
Due to the analyticity in the perturbing parameter $\varepsilon$ 
of the function $u$, we
can expand $u$ in Taylor series around $\varepsilon=0$ as 
$$
u(\theta)=\sum_{n=1}^\infty u_n(\theta)\varepsilon^n\ , 
\eqno{(3.4)} 
$$
for suitable $p$--periodic functions $u_n(\theta)$. Let us expand 
the terms $u_n(\theta)$ in Fourier series as 
$$
u_n(\theta)\ =\ \sum_k\hat u_k^{(n)}\ e^{ik{\theta\over p}}\ , 
$$
where the coefficients $\hat u_k^{(n)}$ (and therefore $u_n(\theta)$) can be
constructed explicitely as follows. Write the function $f_x(x,t)$ as 
$$
f_x(x,t)\ =\ i\ \sum_{m\in{\cal M}}\hat f_m\ (e^{i(2x-mt)}-e^{-i(2x-mt)})\ , 
\eqno{(3.5)} 
$$
where in the case of eq. (2.3) the set $\cal M$ consists of the integers
1,2,...,7 and the $\hat f_m$ can be easily identified by comparison with eq. (2.3). 
From eq. $(3.5)$ it follows that 
$$
f_x(\theta+u(\theta),{\theta\over \omega})\ =\ i\sum_{k=\pm1,m\in{\cal M}} k 
\hat f_m e^{ik(2\theta+2u(\theta)-m\theta{q\over p})}\ . 
$$
Define a power series for the exponential as 
$$
e^{ik(2\theta+2u(\theta)-m\theta{q\over p})}\ \equiv\ \sum_{n=0}^\infty
b_{n,k}^{(m)}(\theta)\varepsilon^n\ . 
$$
It can be easily verified that the coefficients $b_{n,k}^{(m)}$ satisfy the 
following recursive relations: 
$$\eqalign{ 
b_{0,k}^{(m)}&=e^{ik(2\theta-m\theta{q\over p})}\cr 
b_{n,k}^{(m)}&={{2ik}\over n}\ \sum_{h=1}^n h u_h(\theta) b_{n-h,k}^{(m)}\ , 
\qquad n\geq 1\ .\cr}  
$$
Inserting $(3.4)$ in $(3.3)$ one gets 
$$
\sum_{n=1}^\infty D^2u_n(\theta)\varepsilon^n\ =\
i\sum_{n=1}^\infty\varepsilon^n 
\sum_{k=\pm1,m\in{\cal M}}  k\hat f_m b_{n-1,k}^{(m)}\ . 
\eqno{((3.6)}
$$
Moreover, since an explicit computation of the generic term in the summation at
the l.h.s. of $(3.6)$ provides 
$$
D^2u_n(\theta)\ =\ -{1\over {q^2}}\ \sum_k k^2\hat u_k^{(n)} e^{{{ik\theta}\over
p}}\ , 
$$
a comparison of the above relation with the r.h.s. of $(3.6)$ yields the explicit
expression of the coefficients $\hat u_k^{(n)}$. 

\vskip.2in 
\noindent 
Recasting the above formulae, the function $u(\theta)$ can be written as 
$$
u(\theta)=\sum_{n=1}^\infty u_n(\theta)\varepsilon^n\ ,\qquad\qquad 
{\rm with}\qquad u_n(\theta)=\sum_k\hat u_k^{(n)}\sin(n_k{\theta\over p})\ , 
$$
for suitable Fourier indexes $n_k$; the fact that $u_n(\theta)$ is a sum 
of sines is due to the specific form of eq. (2.3) which contains only 
sine--terms. 

\vskip.2in 
\noindent 
As a measure of the regularity of the
function $u(\theta)$, we compute for any $\theta=\theta_0$ the radius of
convergence of the Taylor series as 
$$
\varepsilon_\rho({p\over q})\ \equiv\ \lim_{j\to\infty}
{1\over {(u_j(\theta_0))^{1\over j}}} 
\eqno{(3.7)} 
$$
(which in fact seems not to depend on the choice of $\theta_0$). 
The radius of convergence provides an indication of the stability of the
periodic orbits, showing an approximate value of the perturbing parameter
$\varepsilon=\varepsilon_\rho({p\over q})$ at which the transition from elliptic to
hyperbolic periodic orbits takes place. A numerical investigation of the phase space
suggests that as $\varepsilon$ grows the periodic orbit is surrounded by 
librational curves with increasing amplitude, until they leave place to a
chaotic regime as $\varepsilon$ approaches $\varepsilon_\rho({p\over q})$. 
We make use of this property in order to study the behaviour of the periodic 
orbits as the frequency $p\over q$ is varied.

\vglue2cm 

\noindent 
\bf \S4. RESULTS AND CONCLUSIONS\rm 

\vskip.1in\noindent
According to standard evolutionary theories, satellites and planets were rotating fast
in the past; a constant decrease of the period of rotation about the spin--axis was
provoked by tidal friction due to the internal non--rigidity. Therefore, a common
scenario suggests that celestial bodies were slowed down until they reached their
actual dynamical configuration, being typically trapped in a 1:1 or 3:2 (for Mercury
alone) resonance. This hypothesis implies that higher order resonances (i.e., 2:1,
5:4, 7:3, etc.) were bypassed during the slowing process. However there is actually no
convincing 
explanation concerning the mechanism of \sl escape \rm or \sl capture \rm into a
spin--orbit resonance. In this section we argue that there is a greater probability of
capture into the synchronous or 3:2 resonance. More precisely, we compute an
approximate value of the radius of convergence of the parametric representation of
periodic orbits which provides,  as remarked in \S3, a measure of their stability. We
let the eccentricity vary in a reasonable (astronomical) range of values. In
particular, we consider the following set of periodic orbits with frequencies $p\over
q$ where 
$$
q=1,...,14,\qquad\qquad p=q+1,...,15\ , 
$$
and
$$
p=1,...,14,\qquad\qquad q=p+1,...,15\ . 
$$
In order to have a better precision around the main resonances we consider also the
periodic orbits with frequencies 
$$\eqalign{ 
{1\over 2}&\pm{1\over{10\cdot k}}\ , \qquad\qquad1\pm{1\over{10\cdot k}} \cr
{5\over 4}&\pm{1\over{10\cdot k}}\ , \qquad\qquad{4\over 3}\pm{1\over{10\cdot k}} \cr
{3\over 2}&\pm{1\over{10\cdot k}}\ , \qquad\qquad{5\over 3}\pm{1\over{10\cdot k}} \cr
{7\over 4}&\pm{1\over{10\cdot k}}\ , \qquad\qquad2\pm{1\over{10\cdot k}}\ , \cr} 
$$
where $k=1,...,10$. We computed the quantity ${1\over {(u_j(\theta_0))^{1\over j}}}$ 
for $j=1,...,30$, providing a good approximation to the radius of convergence. The
choice of the point $\theta=\theta_0$ does not influence the result, since the radius
of convergence seems to be generally independent on the initial condition. 

\vskip.1in 
\noindent 
As an example we report in Figure 3 the radius of convergence
$\varepsilon_\rho({1\over 1})$ as a function of the number of iterations 
$j=1,...,30$ (in this case $\theta_0$ was set equal to 1.56). 

\vskip.1in 
\noindent 
For a fixed value of the eccentricity we compute $\varepsilon_\rho({p\over q})$ 
corresponding to the set of rational numbers $p\over q$ listed above. Figures 4 and 5 
show the graphs of $\varepsilon_\rho({p\over q})$ versus $p\over q$ for, 
respectively, the Moon's eccentricity (i.e., $e=0.0549$) and Mercury's eccentricity 
($e=0.2056$). In order to have a more detailed inspection of the behaviour around the
resonances of astronomical interest, we report in panel $a)$ the results around the
1:2 resonance, while panel $b)$ corresponds to the synchronous resonance, panel $c)$ 
to the 3:2 resonance and panel $d)$ to the 2:1 resonance. 

\vskip.1in 
\noindent 
A comparison between Figures 4 and 5 suggests that locally the 1:1 and 3:2 resonances
have a higher probability of capture. In fact, the periodic orbits around these
commensurabilities have very low radii of convergence indicating that the most stable
orbits correspond to the periods ${p\over q}={1\over 1}$ or 
${p\over q}={3\over 2}$. Moreover, comparing Figure $4b$, $4c$ with Figure $5b$, $5c$ 
we notice that there is a slightly preference for the Moon to end--up in the
synchronous resonance, while Mercury might be trapped in the 1:1 or 3:2 resonance, in
agreement with the astronomical observations. 

\noindent 
This discussion leads to the question of the existence of attracting tori
for a \sl dissipative \rm system (i.e., including tidal forces) corresponding to the
most stable resonances. We plan to address this problem in a later study. 

\vskip.2in 
\noindent 
Next we analyze the behaviour of the stability of periodic orbits for different values
of the eccentricity. Figure 6 shows $\varepsilon_\rho({p\over q})$ versus $p\over q$ 
for $e=0.001$ (Figure $6a$), $e=0.01$ (Figure $6b$), $e=0.06$ (Figure $6c$), 
$e=0.2$ (Figure $6d$). For low values of the eccentricity (Figure $6a$) there is only
one marked resonance corresponding to the synchronous commensurability. Moreover, the
amplitude of the curve around the 1:1 resonance provides a measure of the size of the
region of librational motion surrounding the resonance. For values of $p\over q$
bigger than $5\over 2$, the quantity $\varepsilon_\rho({p\over q})$ increases
indefinitely with $p\over q$.  As $e=0.01$ (Figure $6b$), 
beside the 1:1 resonance there appears the 3:2, while the 1:2 and 2:1 resonances are
already present, but with very small amplitudes of librational motion, which become 
meaningful as $e=0.06$ (Figure $6c$). For this value of the eccentricity we have the
complete set of main resonances, i.e. 1:2, 1:1, 3:2, 2:1. Minor resonances appear 
as $e$ is increased up to $e=0.2$ (Figure $6d$), where the 5:4 and 7:4 resonances
become evident. 

\vskip.1in 
\noindent 
These results indicate the existence of a strict relation between the value of the
eccentricity and the birth of resonances. Satellites with low eccentricity 
are suitable candidates for ending--up in the synchronous resonance. However, even
when different resonances arise, the 1:1 periodic orbit seems the most likely final
state due to the amplitude of librational regime around it. When the eccentricity is
highly increased, several new resonances appear. In particular, for Mercury's
eccentricity  (compare with Figure $6d$) different fates are possible, including the
3:2 resonance. 

\vskip.1in 
\noindent 
As a final remark, we suggest that a further development of this study would include 
the effect of dissipative terms and their role in driving satellites to select their 
ending states.

\vfill\eject

\centerline{\bf REFERENCES}

\vglue1cm 

\noindent
A.Cayley, Tables of the developments of functions in the 
theory of elliptic motion, Mem. Roy. Astron. Soc. 29 (1859), 191. 

\vskip.1in
\noindent
A.Celletti, Analysis of resonances in the spin--orbit 
problem in Celestial Mechanics: The synchronous resonance (Part I),  
J. of Appl. Math. and Phys. (ZAMP) 41 (1990), 174. 

\vskip.1in
\noindent
A.Celletti, Construction of librational invariant tori in the 
spin--orbit problem, J. of Applied Math. and Physics (ZAMP) 45 (1994), 61. 

\vskip.1in
\noindent
A.Celletti, L. Chierchia, Construction of stable periodic
orbits for the spin--orbit problem of Celestial Mechanics,
submitted to "Regular and Chaotic Dynamics", Preprint 1998.  

\vskip.1in
\noindent
A.Celletti, C. Falcolini, Construction of invariant tori for the
spin--orbit problem in the Mercury--Sun system,
Celestial Mechanics and Dynamical Astronomy 53 (1992), 113. 

\vskip.1in
\noindent 
P.Goldreich, S.Peale, Spin--orbit coupling in the solar system, 
Astron J. 71 (1966), 425. 

\vskip.1in
\noindent  
P.Goldreich, S.Peale, The dynamics of planetary rotations, 
Ann. Rev. Astron. Astroph. 6 (1970), 287. 

\vskip.1in
\noindent 
J.Henrard, Spin--orbit resonance and the adiabatic invariant, in 
S.Ferraz--Mello, W.Sessin eds., \sl Resonances in the Motion of 
Planets, Satellites and Asteroids, \rm Sao Paulo (1985), 19.

\vskip.1in
\noindent 
J.A.Murdock, Some mathematical aspects of spin--orbit resonance I, 
Cel. Mech. 18 (1978), 237. 

\vskip.1in
\noindent 
S.J.Peale, Rotation of solid bodies in the solar system, Rev. Geoph. and 
Space Physics 11 (1973), 767.

\vskip.1in
\noindent 
J.Wisdom, Rotational dynamics of irregularly shaped satellites, 
Astron. J. 94 (1987), 1350.

\vskip.1in
\noindent
(no author listed) (1997)  {\it The Astronomical Almanac}. 
Washington:U.S. Government  Printing Office.

\vfill\eject 

\bf FIGURE CAPTIONS 

\vglue2cm

\noindent
\bf Figure 1: \rm The spin--orbit geometry. 

\vskip.2in \noindent 
\bf Figure 2: \rm Poincar\'e map associated to eq. (2.3) around the synchronous
resonance in the $(x,\dot x)$--plane 
for $e=0.004$, $\varepsilon=0.2$. The initial conditions are
$(x,y)=(0,0.8)$, $(0,1.5)$, $(0,1)$, $(0,1.1)$, $(1.8,1)$, $(2.1,1)$, $(1.57,1)$. 
Notice that eq. (2.3) is $\pi$--periodic in $x$. 
\sl Librational \rm and \sl rotational \rm
regions are divided by the \sl chaotic separatrix. \rm 

\vskip.2in \noindent 
\bf Figure 3: \rm The radius of convergence $\varepsilon_\rho({1\over 1})$ vs. the 
degree of approximation of the limit $(3.7)$, $j=1,...,30$. Here $\theta_0=1.56$ and 
$e=0.0549$.

\vskip.2in \noindent 
\bf Figure 4: \rm The radius of convergence $\varepsilon_\rho({p\over q})$ vs. $p\over
q$  for $e=0.0549$ (i.e., the eccentricity of the Moon). The dots correspond to the
actual computations associated to the frequencies listed in the text. Lines are due to
interpolation performed by the graphic program. In abscissa are reported 
the rotation numbers $p\over q$ (the 1:1 resonance corresponds to 1, the 1:2 to 0.5
and so on). 

\noindent 
$a)$ 1:2 resonance, $b)$ 1:1 resonance, $c)$ 3:2 resonance, $d)$ 2:1 resonance.

\vskip.2in \noindent 
\bf Figure 5: \rm The radius of convergence $\varepsilon_\rho({p\over q})$ vs. $p\over
q$  for $e=0.2056$ (i.e., the eccentricity of Mercury). The dots correspond to the
actual computations associated to the frequencies listed in the text. Lines are due to
interpolation performed by the graphic program. In abscissa are reported 
the rotation numbers $p\over q$ (the 1:1 resonance corresponds to 1, the 1:2 to 0.5
and so on). 

\noindent 
$a)$ 1:2 resonance, $b)$ 1:1 resonance, $c)$ 3:2 resonance, $d)$ 2:1 resonance.

\vskip.2in \noindent 
\bf Figure 6: \rm Plot of the radius of convergence $\varepsilon_\rho({p\over q})$ vs.
$p\over q$ for the frequencies listed in the text. $a)$ $e=0.001$, $b)$ $e=0.01$, 
$c)$ $e=0.06$, $d)$ $e=0.2$.

\vfill\eject

\vglue4cm 

\def\fig1{\picture 300pt by 190pt (fig1 scaled 800)}
\magnification 1200
\centerline{\fig1}

\vfill\eject

\vglue4cm 

\def\fig2{\picture 350pt by 250pt (fig2 scaled 800)}
\magnification 1200
\centerline{\fig2}

\vfill\eject

\vglue4cm 

\def\fig3{\picture 300pt by 190pt (fig3 scaled 800)}
\magnification 1200
\centerline{\fig3}

\vfill\eject

\vglue4cm 

\def\fig4{\picture 480pt by 190pt (fig4 scaled 650)}
\magnification 1200
\centerline{\fig4}

\vfill\eject

\vglue4cm 

\def\fig5{\picture 480pt by 190pt (fig5 scaled 650)}
\magnification 1200
\centerline{\fig5}

\vfill\eject

\vglue4cm 

\def\fig6{\picture 450pt by 400pt (fig6 scaled 900)}
\magnification 1200
\centerline{\fig6}

\bye